\documentclass[aps,preprint,epsfig,rotate]{revtex4}
\usepackage{graphicx}
\usepackage{bm}
\usepackage{epsfig}

\input epsf
\begin{document}
\title{On the matrix factorization of many-particle Coulomb Hamiltonians}

\author{Alexei M. Frolov}
\email[E--mail address: ]{alex1975frol@gmail.com ; afrolov@uwo.ca}


\affiliation{Department of Applied Mathematics \\
       University of Western Ontario, London, Ontario N6H 5B7, Canada} 

\date{\today}

\begin{abstract}

It is shown that the Coulomb many-particle Hamiltonians are always factorized. This fact can be used to obtain the closed analytical formula(s) for 
the bound state spectra of an arbitrary many-particle Coulomb system. For few- and many-electron atoms and ions these formulas are similar in some sense 
to the Bohr's formula which describes the bound state spectra of the hydrogen atom. 

\noindent 
PACS number(s): 31.30.Gs, 31.15.vj and 32.15.Fn

\end{abstract}

\maketitle
\newpage

\section{Introduction}

In this short communication we discuss the bound state spectra of the actual atoms and/or ions, i.e. one-center Coulomb systems which contain a number of  bound 
electrons. Let us consider the atom/ion which contains $N_e$ bound electrons. In hyperspherical coordinates \cite{Fock}, \cite{Knirk} the Hamiltonian of such an 
atom is written in the form (see \cite{Fro1} for more detail and references)
\begin{eqnarray}
  H(r, \Omega) =  -\frac{1}{2} \Bigl[ \frac{\partial^2}{\partial r^2} + \frac{3 N_e - 1}{r} \frac{\partial}{\partial r} - \frac{\Lambda^2_{N_e}(\Omega)}{r^{2}} 
 \Bigr]  + \frac{W(\Omega)}{r} \; \; \; \label{HHH}
\end{eqnarray}
where $\Lambda^2_{N_e}(\Omega)$ is the hypermomentum of the atom, while $W(\Omega)$ is the hyperangular part of the Coulomb interaction potential which includes
electron-nucleus and electron-electron parts. Here and everywhere below we apply the atomic units (where $\hbar = 1, \mid e \mid = 1$ and $m_e = 1$) and use the notation 
defined in \cite{Fro1}. In particular, $\Omega$ means the $3 N_e - 1$ angular and hyperangular electron's coordinates (compact variables), while $r$ designates the 
hyper-radius. In the basis of the `physical' hyperspherical harmonics \cite{Fro1} ${\cal Y}_{\vec{K}(c),\vec{\ell}(c),\vec{m}(c)}(\Omega)$ (below HH, for short) and for 
the radial functions represented in the form $r^{-\frac{3 N_e - 1}{2}} \Psi(r)$ this Hamiltonian takes the self-conjugate form
\begin{eqnarray}
  H(r) = \frac{1}{2} \Bigl[ p^{2}_r + \frac{\Bigl(\hat{K} + \frac{3 N_e + 1}{2}\Bigr) \Bigl(\hat{K} + \frac{3 N_e - 1}{2} - 1\Bigl)}{r^{2}} \Bigr] + \frac{\hat{W}}{r} 
  \; \; \; \label{HHHa}
\end{eqnarray}
where the hyper-radial momentum operator $p_r$ is defined as follows $p_r = (-\imath) \frac{\partial}{\partial r}$, $\hat{W}$ is the matrix of the hyperangular part of 
the Coulomb interaction potential in the basis of physical HH (definition of the physical HH can be found, e.g., in \cite{Fro1}, \cite{Our1}). Also, in Eq.(\ref{HHHa}) 
the notation $\hat{K}$ stands for the matrix of hypermomentum which is a diagonal matrix in the basis of hyperspherical harmonics. The self-conjugate form of these two 
operators is more appropriate for our present purposes. 

In \cite{Fro1} (see also \cite{Fro1986}) we have shown that the atomic Hamiltonian $H(r)$, Eq.(\ref{HHHa}), is factorized, i.e. $H(r)$ is represented in the form
\begin{eqnarray}
  H = \Theta^{\ast}_1(r) \Theta_1(r) + \hat{a}_1 \; \; \; \label{factHa}
\end{eqnarray} 
where $\hat{a}_1$ is a matrix defined below, while the operator $\Theta_1(r)$ and its adjoint operator $\Theta^{\ast}_1(r)$ are the first-order differential operators 
defined as follows
\begin{equation}
 \Theta_1(r) = \frac{1}{\sqrt{2}} \Bigl[ -\imath p_r + \frac{\hat{\beta}_1}{r} + \hat{\alpha}_1 \Bigl] = 
 \frac{1}{\sqrt{2}} \Bigl[ -\frac{\partial}{\partial r} + \frac{\hat{\beta}_1}{r} + \hat{\alpha}_1 \Bigl] \; \; \; \label{Thm1}
\end{equation}
and  
\begin{equation}
 \Theta^{\ast}_1(r)  = \frac{1}{\sqrt{2}} \Bigl[ \imath p_r + \frac{\hat{\beta}_1}{r} + \hat{\alpha}_1 \Bigl]
 = \frac{1}{\sqrt{2}} \Bigl[ \frac{\partial}{\partial r} + \frac{\hat{\beta}_1}{r} + \hat{\alpha}_1 \Bigl] \; \; \; \label{Thm1c}
\end{equation}
where the notations $\hat{\beta}_1, \hat{\alpha}_1$ and $\hat{a}_1$ from Eq.(\ref{factHa}) stand for the symmetric, infinite-dimensional, in principle, matrices which 
do not commute with each other. In actual applications the dimensions of these matrices coincide with the total number of hyperspherical harmonics used. By substituting 
these two expressions, Eqs.(\ref{Thm1}) - (\ref{Thm1c}), into Eq.(\ref{factHa}) one finds the following equations for the $\hat{\alpha}_1, \hat{\beta}_1$ and 
$\hat{a}_1$ matrices:
\begin{eqnarray}
    & & \hat{\beta}_1 (\hat{\beta}_1 - 1) = \Bigr(\hat{K} + \frac{3 N_e - 1}{2}\Bigl) \Bigr(\hat{K} + \frac{3 N_e - 1}{2} - 1\Bigl)  \; \; \; \label{eqx} \\
    & & \hat{\alpha}_1 \hat{\beta}_1 + \hat{\beta}_1 \hat{\alpha}_1 = 2 \hat{W} \; \; \; \label{eqy} \\
    & & \hat{a}_1 = -\frac12 \hat{\alpha}^2_1 \; \; \; \label{eqz}
\end{eqnarray}
where the matrix of hypermomentum $\hat{K}$ is a diagonal matrix in the basis of hyperspherical harmonics (or, in $K-$representation, for short). Solution of 
Eq.(\ref{eqx}) is written in the form 
\begin{equation}
    \hat{\beta}_1 = \hat{K} + \frac{3 N_e - 1}{2}   \; \; \; \label{eqxx}
\end{equation}
where we use the fact that the atomic wave function must be regular at $r = 0$, i.e. at the atomic nucleus. As follows from this equation the matrix $\hat{\beta}_1$ is 
also diagonal in $K-$representation.  Below, we apply only this $K-$representation, since it substantially simplifies a large number of formulas derived below. In 
particular, by using the formula from \cite{Belman} (see Chapter 10, \$ 18) we can write the explicit expression for the $\hat{\alpha}_1$ matrix
\begin{equation}
    \hat{\alpha}_1 = 2 \int_{0}^{+\infty} \exp(-\hat{\beta}_1 t) \hat{W} \exp(-\hat{\beta}_1 t) dt  \; \; \; \label{eqyy}
\end{equation}
Since the $\hat{\beta}_1$ matrix is diagonal, then for the $(ij)-$matrix element of the $\hat{\alpha}_1$ matrix one finds
\begin{equation}
  \Bigl[ \hat{\alpha}_1 \Bigr]_{ij} = \frac{2 W_{ij}}{[\beta_1]_{ii} + [\beta_1]_{jj}} = \frac{2 W_{ij}}{[\beta_{1}]_{i} + [\beta_{1}]_{j}} = 
  \frac{2 W_{ij}}{K_{i} + K_{j} + 3 N_e - 1}
\end{equation}
Finally, we can determine the $\hat{a}_1$ matrix from Eq.(\ref{eqz}). In particular, for the $(ij)-$matrix elements of the $\hat{a}_1$ matrix we have 
\begin{equation}
  \Bigl[ \hat{a}_1 \Bigr]_{ij} = - 2 \sum_{k} \frac{W_{ik}}{\beta_{i} + \beta_{k}} \cdot \frac{W_{kj}}{\beta_{k} + \beta_{j}} = 
     - 2 \sum_{k} \frac{1}{\beta_{i} + \beta_{k}} \Bigl[ W_{ik} W_{kj} \Bigr] \frac{1}{\beta_{k} + \beta_{j}}  \; \; \; \label{eqzz}
\end{equation}

At the second stage of the procedure, we introduce the radial operators $\Theta_n(r)$ for $n = 2, 3, \ldots$, which are similar to the operators $\Theta_1(r)$ defined 
above (see, Eq.(\ref{Thm1}), i.e.
\begin{equation}
 \Theta_n(r) = \frac{1}{\sqrt{2}} \Bigl[ -\imath p_r + \frac{\hat{\beta}_n}{r} + \hat{\alpha}_n \Bigl]
 = \frac{1}{\sqrt{2}} \Bigl[ - \frac{\partial}{\partial r} + \frac{\hat{\beta}_n}{r} + \hat{\alpha}_n \Bigl] \; \; \; \label{Thmn}
\end{equation}
The adjoint operators take the form 
\begin{equation}
 \Theta^{\ast}_n(r) = \frac{1}{\sqrt{2}} \Bigl[ \imath p_r + \frac{\hat{\beta}_n}{r} + \hat{\alpha}_n \Bigl]
 = \frac{1}{\sqrt{2}} \Bigl[ \frac{\partial}{\partial r} + \frac{\hat{\beta}_n}{r} + \hat{\alpha}_n \Bigl] \; \; \; \label{Thmnc}
\end{equation}
The logically closed method of matrix factorization method is based on the following `ladder' conditions (see, e.g., \cite{Green}) 
\begin{eqnarray}
  \Theta_n(r) \Theta^{\ast}_n(r) + \hat{a}_n = H_{n+1} = \Theta^{\ast}_{n+1}(r) \Theta_{n+1}(r) + \hat{a}_{n+1} \; \; \; \label{factHmn}
\end{eqnarray}
which must be obeyed for $n = 1, 2, \ldots$. By substituing the explicit expressions, Eqs.(\ref{Thmn}) and (\ref{Thmnc}) into Eq.(\ref{factHmn}) we obtain the following 
equations for the $\hat{\beta}_n, \hat{\beta}_{n+1}, \hat{\alpha}_{n}, \hat{\alpha}_{n+1}, \hat{a}_n$ and $\hat{a}_{n+1}$ matrices
\begin{eqnarray}
 & &  \hat{\beta}_{n+1} (\hat{\beta}_{n+1} - 1) = \hat{\beta}_{n} (\hat{\beta}_{n} + 1) \; \; \; , \; \; \; \label{I1} \\ 
 & & \hat{\alpha}_{n} \hat{\beta}_{n} + \hat{\beta}_{n} \hat{\alpha}_{n} = 2 \hat{W} = \alpha_{n+1} \hat{\beta}_{n+1} + \hat{\beta}_{n+1} \hat{\alpha}_{n+1} 
  \; \; \; , \; \; \; \label{I2} \\
 & & \hat{a}_{n} = -\frac12 \alpha^{2}_{n} \; \; \; , \; \; \; \hat{a}_{n+1} = -\frac12 \alpha^{2}_{n+1} \; \; \;   \label{I3}
\end{eqnarray} 
These matrix equations look very similar to the analogous numerical equations known in the traditional (or numerical) factorization method for the hydrogen-like atomic 
systems (see, e.g., \cite{Green}). However, Eqs.(\ref{I1}) - (\ref{I3}) are written for the symmetric, infinite-dimensional matrices, which do not commute 
with each other, e.g., the $\hat{\beta}_n$ matrix do not commute with the $\hat{\alpha}_{n}$ and $\hat{a}_{n+1}$ matrices, etc. Solution of these equations, 
Eqs.(\ref{I1}) - (\ref{I3}), regular at $r = 0$ is written in the form
\begin{eqnarray}
 & & \hat{\beta}_{n+1} = \hat{\beta}_{n} + 1 = \ldots = \hat{\beta}_1 + n = \hat{K} + \frac{3 N_e - 1}{2} + n \; \; \; \label{eqxxm} \\
 & & \hat{\alpha}_{n+1} = 2 \int_{0}^{+\infty} \exp(-\hat{\beta}_{n+1} t) \hat{W} \exp(-\hat{\beta}_{n+1} t) dt  \; \; \; \label{eqyym} \\
 & & \hat{a}_{n+1} = -\frac12 \alpha^{2}_{n+1} \; \; \; \label{eqzzm}
\end{eqnarray}

The equation, Eq.(\ref{eqyym})), produces the explicit formula for the $(ij)-$matrix element of the $\hat{\alpha}_{n+1}$ matrix
\begin{equation}
  \Bigl[ \hat{\alpha}_{n+1} \Bigr]_{ij} = \frac{2 W_{ij}}{[\beta_{n+1}]_{ii} + [\beta_{n+1}]_{jj}} = \frac{2 W_{ij}}{[\beta_{1}]_{i} + [\beta_{1}]_{j} + 2 n} = 
  \frac{2 W_{ij}}{K_{i} + K_{j} + 3 N_e - 1 + 2 n}\; \;  \;
  \label{fres1}
\end{equation}
where $[\beta_{1}]_{i}$ is the $(ii)-$matrix element of the diagonal $\hat{\beta}_{1}$ matrix and we can write in the general case that $[\beta_{n+1}]_{ij} = \delta_{ij} 
[\beta_{n+1}]_{ii} =\delta_{ij} [\beta_{n+1}]_{i}$ and $[\beta_{1}]_{ij} = \delta_{ij} [\beta_{1}]_{ii} = \delta_{ij} [\beta_{1}]_{i}$. This leads to the following 
analytical expression for the $(ij)-$matrix elements of the $\hat{a}_{n+1}$ matrix
\begin{eqnarray} 
 & & \Bigl[ \hat{a}_{n+1} \Bigr]_{ij} = - 2 \sum_{k} \frac{W_{ik}}{[\beta_{1}]_{i} + [\beta_{1}]_{k} + 2 n} \cdot \frac{W_{kj}}{[\beta_1]_{k} + [\beta_1]_{j} + 2 n} 
   \; \; \; \label{eqzzm1} \\
 &=& - 2 \sum_{k} \frac{1}{K_{i} + K_{k} + 2 n + 3 N_e - 1} \Bigl[ W_{ik} W_{kj} \Bigr] \frac{1}{K_{k} + K_{j} + 2 n + 3 N_e - 1} \nonumber
\end{eqnarray}
where $K_{i}$ are the matrix elements of the diagonal $\hat{K}$-matrix (the matrix of hypermomentum) and $n \ge 0$, where $n$ is the hyper-radial quantum number which is 
always integer and non-negative. Formally, this formula is a direct generalization of the Bohr's formula, originally derived by N. Bohr (in 1913) for the hydrogen atom, 
to an atom/ion which contains $N_e$ bound electrons. For $N_e = 1$ the formula Eq.(\ref{eqzzm1}) exactly coincides with the Bohr's formula (in atomic units). Indeed, in 
this case $3 N_e - 1 = 2$, $\hat{W}_{ij} = -Q \delta_{ij}, K_i = K_j = \ell$ and $\ell$ is the good quantum number. Therefore, one finds from Eq.(\ref{eqzzm1}) $E_i = 
\Bigl[ \hat{a}_{n+1} \Bigr]_{ii} = - \frac{Q^2}{2 (\ell + 1 + n)^2}$. 
 
Note that for the one-electron atom/ion the energy bound state spectrum can be determined without any reference to the wave functions. However, this is not the case for 
few- and many-electron atoms/ions which have been analyzed in \cite{Fro1}. Here we present a very brief description of the results derived in \cite{Fro1}. For an arbitrary
atom/ion with $N_e$ bound electrons we chose some atomic term $\Bigl[ L, M, S, S_z, \pi \Bigr]$ \cite{Fro1}. For this term we construct the system of physical HH 
(hyperangular basis). In these basis calculate we calculate all elements of the atomic matrix of the Coulomb potential $\hat{W}$ (see, e.g., \ref{HHH} and \ref{HHHa} above). 
By using the known expressions for the matrix elements of the $\hat{W}$ matrix we can determine the matrix elements of the following matrices $\hat{{\cal A}}(n)$ \cite{Fro1}
\begin{eqnarray}
  [\hat{{\cal A}}(n)]_{ij} &=& \frac{\Gamma(K_i + K_j + 2 n + 3 N_e)}{\sqrt{\Gamma(2 K_i + 2 n + 3 N_e) \Gamma(2 K_j + 2 n + 3 N_e)}} \cdot 
  \frac{2 W_{ij}}{K_{i} + K_{j} + 3 N_e - 1 + 2 n} \; \; \label{wf158a} \\
 &=& \frac{(K_i + K_j + 2 n + 3 N_e - 1)!}{\sqrt{(2 K_i + 2 n + 3 N_e - 1)! (2 K_j + 2 n + 3 N_e - 1)!}} \cdot \frac{2 W_{ij}}{K_{i} + K_{j} + 3 N_e - 1 + 2 n} \nonumber  
\end{eqnarray}
where $n = 0, 1, 2, \ldots$. The matrices $[\hat{{\cal A}}(n)]$ are symmetric and all their eigenvalues are negative. At the second stage of the procedure we determine the 
lowest eigenvalue $\lambda_{n+1}$ of each of these matrices $\hat{{\cal A}}(n)$, where $n = 0, 1, 2, \ldots$. The total energies $E_{n+1}$ of the corresponding bound states 
in the atom/ion with $N_e$ bound electrons are simply related to the $\lambda_{n+1}$ eigenvalues by the formula $E_{n+1} = -\frac12 \lambda^{2}_{n+1}$. This gives us the 
complete energy spectrum of bound states for this atomic term $\Bigl[ L, M, S, S_z, \pi \Bigr]$. The procedure to obtaining the corresponding eigenfunctions is described in 
\cite{Fro1}. An obvious advantage of our approach follows from the fact that all bound state energies are determined in a closed analytical form as the solutions of simply 
related eigenvalue problems. This allows one to investigate explicitly the dependencies of the total energies of one atomic term upon the conserving quantum numbers (e.g., 
upon the hyper-radial quantum nuber $n$ which is also called the number of excitations). Also, we do not need to solve any hyper-radial eigenvalue problem. In other words, 
the hyper-radial dependence of the actual wave function of the $N_e-$electron atom/ion is uniformly determined by the corresponding hyperangular matrix of the potenetial 
energy. There are some other advantages of our procedure, but we cannot discuss them in our short communication. 

In conclusion we want to answer the following question: why is it possible to determine the bound state spectra of atoms and ions only in hyperspherical coordinates? In 
other words, why we cannot use for these purposes any alternative set of coordinates, e.g., Cartesian coordinates, or any other set of coordinates which include two or more 
non-compact variables? After careful investigation of this problem and after reading of \cite{Fock}, \cite{Fock1} and \cite{Elut} I understood the reason of such a special 
role of hyperspherical coordinates. It can be formulated in the form: in hyperspherical coordinates the bound and continuous parts of atomic spectra are separatred by a 
simple algebraic (even arithmetic) transformation of the wave function, while in other coordinates mentioned above it is not possible to perform. This also follows from the 
following theorem \cite{FroFirst}: the three operators $S, T$ and $U$ defined by the formulas
\begin{eqnarray}
  S = \frac12 r \Bigl( p_{r}^2 + \frac{ K^{2}_{N_e}(\Omega)}{r^2} + 1 \Bigr) \; \; \; , \; \; \; T = r p_r \; \; \; and \; \; \; 
  U = \frac12 r \Bigl( p_{r}^2 + \frac{ K^{2}_{N_e}(\Omega)}{r^2} - 1 \Bigr) \; \; \; \label{AFrol} 
\end{eqnarray}
form the $O(2,1)-$algebra, i.e. the obey the following commutation relations
\begin{eqnarray}
  [ S, T ] = - \imath U \; \; \; , \; \; \; [ T, U ] = \imath S \; \; \; , \; \; \; [ U, S ] = - \imath T \; \; \; \label{AFrol1} 
\end{eqnarray}
and the Casimir operator of the second order ($C_2$) equals 
\begin{equation}
  C_2 =  K^{2}_{N_e}(\Omega) = \Lambda^2_{N_e}(\Omega) + \Bigl(\frac{3 N_e - 1}{2}\Bigr)^2 - \Bigl(\frac{3 N_e - 1}{2}\Bigr) \label{Kasimir}
\end{equation}
All notations used in Eqs.(\ref{AFrol}) - (\ref{Kasimir}) are exactly the same as in Eq.(\ref{HHH}). The proof of this theorem can be found in \cite{FroFirst}. Based on this 
theorem one can reproduce all our results obtained in \cite{Fro1} and in this study (see, e.g., \cite{FroFirst}). The coincidence of the Casimir operator ($C_2$) of this 
hyper-radial $O(2,1)$ algebra with the analogous Casimir operator of the hyperangular (compact) $O(3N_e)$-algebra is not a random fact \cite{Mosh}.

\end{document}